\documentclass[overfull,figures]{epl}
\usepackage{amssymb,amsmath}
\usepackage{epsf,colordvi,graphicx,color,amsbsy,amssymb}
\usepackage[T1]{fontenc}

\title{Collision statistics in a dilute granular gas fluidized by vibrations in low gravity}
\author{E. Falcon\inst{1}\thanks{E-mail: \email{Eric.Falcon@ens-lyon.fr}, URL: http://perso.ens-lyon.fr/eric.falcon/} \and S. Auma\^\i tre\inst{2} \and P. \'Evesque\inst{3} \and F. Palencia\inst{4} \and C. Lecoutre-Chabot\inst{4} \and S. Fauve\inst{2} \and D. Beysens\inst{4} \and  Y. Garrabos\inst{4}}
\institute{
\inst{1} Laboratoire de Physique, ENS Lyon, UMR 5672 CNRS, 46, all\'ee d'Italie, 69 007 Lyon, France\\
\inst{2} LPS, ENS Paris, UMR 8550 CNRS, 24, rue Lhomond, 75 005 Paris, France\\
\inst{3} Laboratoire MSSMat, \'Ecole Centrale Paris, UMR 8579 CNRS, 92 295 Chatenay--Malabry, France\\
\inst{4} ESEME - CEA - CNRS, ICMCB, 87 av. Schweitzer, Universit\'e Bordeaux 1, 33 608 Pessac Cedex, France
}
\shortauthor{E. Falcon \etal}
\shorttitle{Collision statistics in a dilute granular gas in low gravity}

\begin{document}
\maketitle

\begin{abstract}
We report an experimental study of a dilute ``gas'' of inelastically colliding particles excited by vibrations in low gravity. We show that recording the collision frequency together with the impulses on a wall of the container gives access to several quantities of interest. We observe that the mean collision frequency does not scale linearly with the number $N$ of particles in the container. This is due to the dissipative nature of the collisions and is also directly related to the non extensive behaviour of  the kinetic energy (the granular temperature is not intensive).   
\end{abstract}

\section{Introduction} Since Maxwell and Boltzmann predictions in 1860-70, it is well-known that molecules of a gas move erratically with a Gaussian velocity distribution, as experimentally verified later \cite{Miller55}. This allows thermodynamic and transport properties of molecular gases to be described.  However, these kinetic theory results do not hold if the particle interactions are dissipative or depend on their velocities (e.g., in relativistic plasmas \cite{Landau}). A well-known example of dissipative gas is the granular gas (see \cite{GranularGas,GranularGasD} for a recent collection of papers). Since collisions between granular particles are inelastic, a continuous input of energy (by vibrating a piston or the container) is required to reach a nonequilibrium steady state. In this regime,  granular matter sometimes seems to behave like an molecular gas in which particles follow erratic motions, but several experiments have displayed striking different properties: Instability of the homogeneous density state leading to cluster formation \cite{Gollub97,FalconMug,Falcon01}, non-Gaussian nature of the velocity distribution \cite{Rouyer00}, anomalous scaling of the pressure \cite{Falcon01,McNamara03}. These effects have been also numerically simulated and some of them have been theoretically understood \cite{Goldhirsch03}.

In this paper, we report a 3D experiment of a dilute granular medium fluidized by sinusoidal vibrations in a low gravity environment. The motivation for low gravity is to achieve an experimental situation in which inelastic collisions are the only interaction mechanism, and where only one ``input'' variable (the inverse vibration frequency) has the dimension of time \cite{FalconMug}. This eliminates possible resonances between the time of flight of a particle under gravity and the period of vibration. The aim is to observe new phenomena which result from the inelasticity of the collisions, thus absent in molecular gases. We first study the scaling of the collision frequency with a container wall with respect to the vibration velocity, $V$, and the particle number, $N$. We also measure the time lag distribution between successive collisions with the wall and the impulse distribution.  We show in particular that two measurements display  significant differences from the behaviours observed in molecular gases: the scalings of the collision frequency and of the particle impulse distribution with $N$. The scaling of the granular temperature with $V$ has been extensively investigated \cite{GranularGas}, but there exists only one 2D experiment for the scaling with $N$ \cite{Warr95}. We emphasize that velocity distributions in granular gases have been measured so far only for nearly 2D geometries. In the 3D case, it is much easier to measure the distribution of impact velocities at a boundary as done here. This measurement involves a similar information content and can be easily compared to molecular dynamics simulations \cite{Aumaitre2}. 
\begin{figure}[h]
\centerline{
\epsfysize=50mm
\epsffile{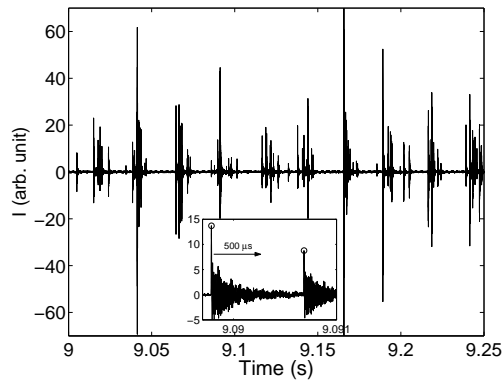} }
\caption{Typical time recording of the force sensor [impulse response $I(t)$] during 10 periods of vibration showing 106 collisions. Inset: zoom of this signal during 1.3 ms showing two detected collision peaks ($\circ$) and the typical damping time of the oscillatory response of the sensor. The parameters of vibration are: $N=12$, $f=40$ Hz, $A=1.96$ mm (not listed in Table\ \ref{tab01}).}
\label{Fig01}
\end{figure}

\section{Experimental setup} A fixed transparent Lexan tube, $D=12.7$ mm in inner diameter and $L=10$ mm in height, is filled with $N$ steel spheres, $d = 2$ mm in diameter. Experiments have been performed with $N=12$, 24, 36 and 48 respectively  corresponding to $n=$ 0.3, 0.6, 0.9 and 1.2 particle layers at rest (packing fraction from 0.04 to 0.18).  A piezoelectric force sensor (PCB 200B02), 12.7 mm in diameter, is fixed at the top of the cell in order to record the particle collisions with the upper wall. A piston made of duralumin, 12 mm in diameter, is driven sinusoidally at the bottom of the cell by an electromagnetic shaker. The frequency $f$ is in the range 40 to 91 Hz and the maximal displacement amplitude $A$ is varied from 0.4 to 2 mm. The vibration parameters during the time line are listed in Table\ \ref{tab01}. Vibration amplitudes are measured by piezoelectric accelerometers (PCB 356A08) screwed in the shaft in a triaxial way. Typical output sensitivities in the vibration direction and in the perpendicular directions are, respectively, 0.1 and 1 V/$g$, where $g=9.81$ m/s$^2$ is the acceleration of gravity.  Typical force sensor characteristics are a 11.4 mV/N output sensitivity, a 70 kHz resonant frequency, and a 10 $\mu$s rise time.  Low gravity environment  (about $\pm 5\times 10^{-2} g$) is repetitively achieved by flying with the specially modified {\em Airbus A300 Zero-G} aircraft through a series of parabolic trajectories which result in low gravity periods, each of 20 s. An absolute acceleration sensor allows the detection of the low gravity phases and the automatic increment of the vibration parameters after each parabola. The output signals of force, respectively accelerations, are stored on a computer on 16 bits at a 2 MHz sampling rate, respectively, on 12 bits at 10 kHz. 

\begin{table}[ht]
\begin{center}
\begin{tabular}{ccccccc} 
$N$  & $A$ & $f$ & $V$ & $\Gamma$ & $N_w$ & Symbols \\
        & (mm)  & (Hz) & (m/s)  & & & in Fig.\ \ref{Fig03} \\
\hline
12 &0.92 & 40 & 0.23 & 5.9 & 2591 & $\times$\\
12 &0.65 & 59.7 & 0.24 & 9.3 & 2605 & $\circ$\\ 
12 &0.88 & 80 & 0.44 & 22.7 & 5756 & $\bullet$\\ 
12 &0.64 & 90.9 & 0.37 & 21.4 & 4617 & $+$\\ 
24 &0.96 & 40 & 0.24 & 6.2 & 5097 & $\ast$\\ 
24 &0.67 & 59.7 & 0.25 & 9.6 & 4078 & $\Diamond$\\ 
24 &0.89 & 80 & 0.44 & 22.8 & 8362 & $\bigtriangledown$\\
36 &0.44 & 40& 0.11 & 2.8 &  2538 & $\bigtriangleup$\\
36 &0.67 & 59.7& 0.25 & 9.7 &  6496 & $penta.$\\
36 &0.89 & 80& 0.44 &  22.8 & 9744 & $\circ$\\
36 &0.69 & 90.9& 0.39 &  22.9 & 9741 & $\times$\\
48 &0.42 & 40& 0.11 &  2.7  &  2728 & $\times$ \\ 
48 &0.69 & 59.7& 0.26 & 9.9 &  8650 & $hexa.$\\ 
48 &0.89 & 80& 0.45 & 22.9&  10906 & $\circ$ \\ 
48 &0.73 & 90.9& 0.41 & 24.2 &  12512 & $\Box$
\end{tabular}
\end{center}
\caption{Vibration parameters during each parabola. $V=2\pi Af$ and $\Gamma=4\pi^{2}Af^{2}/g$ are respectively the maximal piston velocity and the dimensionless acceleration. The number of collisions $N_w$ on the sensor is detected during $\theta = 16$ s of low gravity to avoid transient states.}
\label{tab01}
\end{table}

\section{Detection of collisions} A typical time recording of the force sensor shows a succession of peaks corresponding to particle collisions, as displayed in Fig.\ \ref{Fig01}  for 10 periods of vibration.  Bursts of peaks roughly occur in phase with the vibration but the number of peaks in each burst and their amplitude are random (see Fig.\ \ref{Fig01}). A peak corresponds to the collision of a single sphere, which leads to an almost constant impact duration from 5 to 6 $\mu$s for our range of particle velocities $v$ (assumed of the order of $V$). Indeed, the Hertz's law of contact between a sphere of radius $R$ and a plane made of same material, leads to a duration of collision $\tau = Y R / v^{1/5}$, where $Y=6.9\times 10^{-3}$ $($s$/$m$)^{4/5}$ for steel  \cite{Falcon98}. The signal recorded by the sensor corresponds to an impulse response, $I(t)$. Each peak due to a collision is thus followed by an oscillatory tail at the sensor resonance frequency (roughly 100 kHz) damped over 500 $\mu$s (see inset of Fig.\ \ref{Fig01}). A thresholding technique is applied to the signal in order to detect the collisions. We have to discard a time interval of 100 $\mu$s around each detected peak in order to avoid counting the first maxima of each oscillatory tail as additional collisions. 
Thus an additional weak collision occuring in the oscillatory tail due to the previous one may be missed by our detection process. However, the discarded time interval is small compared to the mean time lag (a few ms) between successive collisions if their statistics is assumed Poissonian (see below). Consequently, the probability of possibly discarded collisions is small. 

\section{Collision frequency scaling} 
\begin{figure}[h]
\centerline{
\epsfysize=60mm
\epsffile{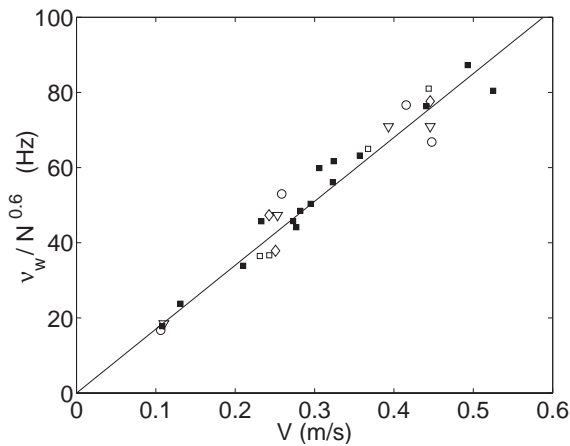}}
\caption{Collision frequency rescaled by the number of particles, $\nu_w /  N^{0.6}$, as a function of $V$ for $N=$12 ($\Box$) and ($\blacksquare$); 24($\Diamond$); 36($\bigtriangledown$); 48($\circ$). $\blacksquare$-marks are from a previous set of experiments at fixed $N=12$ for 15 different velocities which are not listed in Table\ \ref{tab01}. Solid line corresponds to the fit $\nu_w / N^{0.6} = V/l_0$ where $l_0\simeq5.9$ mm.}
\label{Fig02} 
\end{figure}

The number of collisions $N_w$ with the top wall (i.e., the sensor) is obtained by the previous thresholding technique, for each parameter listed in Table\ \ref{tab01}, during $\theta=16$ s of low gravity to avoid possible transient states. For a fixed number of particles, $N$, Fig.\ \ref{Fig02} shows that the collision frequency, $\nu_w = N_w/\theta$, is proportional to the maximal piston velocity, $V$, for $0.1 \le V \le 0.5$ m/s.  As also shown in Fig.\ \ref{Fig02}, $\nu_w \propto VN^{\alpha},$ with $\alpha = 0.6 \pm 0.1$ for our range of $N$. This result strongly differs from the kinetic theory of molecular gases for which $\nu_w$ varies linearly with $N$.
It cannot be explained either in the very dilute limit (Knudsen regime). Indeed, assuming that each particle mostly collides with the boundaries of the container and does not interact with others,  leads to $\nu_w \propto VN / [2(L-d)]$. 
Therefore particles do interact significantly with each other through inelastic collisions. We will show below that this anomalous scaling is a consequence of the dissipative nature of collisions. Note also that this scaling law with $N$ has been recently recovered in 2D numerical simulations \cite{Aumaitre2}. It thus appears to be a robust and generic behaviour of granular gases as we will explain below.

\section{Time lag distribution} 
\begin{figure}[ht]
\centerline{
\epsfysize=78.95mm
\epsffile{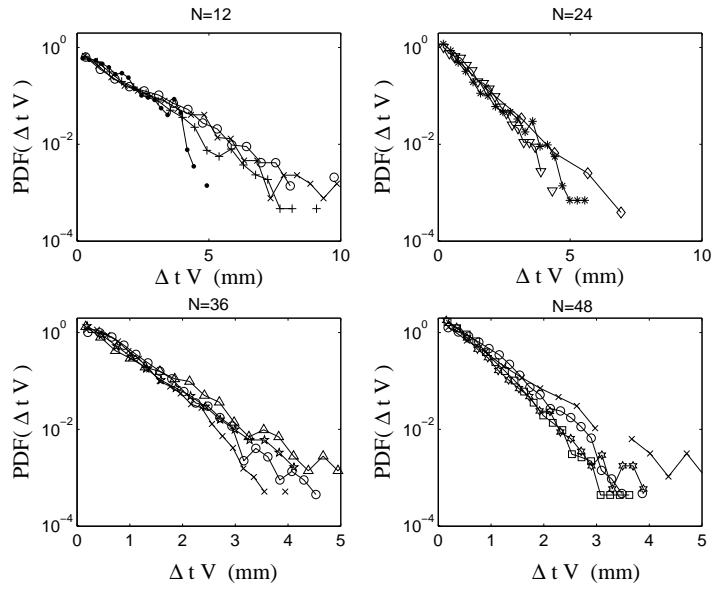} }
\caption{Probability density functions of the time lag $\Delta t$ between two successive collisions rescaled by $V$ ($V = 2\pi Af$), for $N=12$, 24, 36 and 48 particles, and for different vibration parameters (see symbols in Table\ \ref{tab01}).}
\label{Fig03} 
\end{figure}

The probability density functions (PDF) of the time lag $\Delta t$ between two successive collisions with the top wall is displayed in Fig.\ \ref{Fig03} for 4 different values of $N$ and various parameters of vibration. These PDFs are found to decrease exponentially with  $\Delta t$ and to scale like $V$ for our range of $V$. This exponential distribution for the time lag statistics is the expected one for Poissonian statistics. As already shown for the data of Fig.\ \ref{Fig02}, these PDFs can be collapsed by the $N^{0.6}$ rescaling.  We also observe in Fig.\ \ref{Fig03} that even the largest values of  $\Delta t V$ are smaller than $L$. In our range of $N$, the   Knudsen number, $K = l / L$, is in the range $0.1 - 1$, where $l$ is the mean free path, $l = \Omega/(N \pi d^2)$. We are thus in a transition regime from a Knudsen regime to a kinetic regime. It corresponds to a crossover between the very dilute regime for which each particle mostly collides with the boundaries ($l$ of order $L$ independent of $N$), to the kinetic regime ($l$ inversely proportional to $N$). Finally,  if the amplitude of vibration, $A$, is not negligible with respect to $L$ (i.e., $A/L \ge 0.17$), the time lag distributions are no longer exponential (not shown here). 
\begin{figure}[hb]
\centerline{
\epsfysize=78.95mm
\epsffile{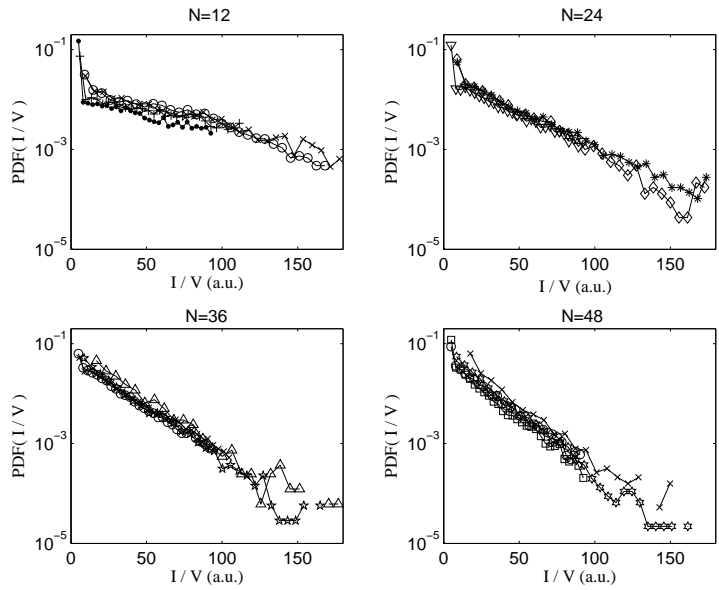} }
\caption{Probability density functions of the impulse $I$ of the impacts on the sensor rescaled by $V$, for different vibration parameters. Symbols  are the same as Fig.\ \ref{Fig03}.} 
\label{Fig04} 
\end{figure} 

\section{Impulse distribution} The PDFs of the maxima $I$ of the impacts recorded in Fig.\ \ref{Fig01} are displayed in Fig.\ \ref{Fig04} for 4 different values of $N$ and various parameters of vibration. Note that the low impulse events are not resolved  because of noise. Indeed, we expect that the PDFs vanish for $I=0$. We first observe that they scale like $V$, for our range of $V$ (see Fig.\ \ref{Fig04}). Second, they display exponential tails with a slope increasing with the particle number $N$. 
Third, the PDF for different values of $N$ can be roughly collapsed when $I$ is scaled like  $V / N^{\beta}$ with $\beta \approx 0.8 \pm 0.2$. This shows that the mean particle velocity $\overline v$ near the wall scales like,
$\overline v \propto V / N^{\beta}$, which gives for the granular temperature near the wall, $T_w \propto V^2 / N^{2 \beta}$ (see below). Finally, we observe that the shape of the distributions of $I$ also differ from kinetic theory of molecular gas. They display an exponential tail instead of the Gaussian one. Note however, that it has been observed many times that a universal shape of the bulk velocity distribution do not exist for granular gases. In particular, a significant effect of side walls has been reported \cite{Rouyer00}. 

\section{Discussion and concluding remarks} The experimental results obtained on the collision frequency and the distribution of impulse at the wall are related and can be used to extract information on various quantities of interest. Indeed, keeping only quantities that depends on $V$ and $N$,  we have $\nu_w \propto \rho_w \overline {v} \propto \rho_w \overline {I}$ where $\rho_w$ is the particle density at the wall. Thus, we get $\rho_w \propto N^{\alpha + \beta}$ and $T_w \propto V^2 N^{- 2\beta}$. The density of particles close to the wall opposite to the piston increases faster than $N$ ($\alpha + \beta = 1.4 \pm 0.3$) because the density gradient becomes larger when $N$ increases. Indeed, it is well known that the granular temperature decreases away from the piston because of inelastic collisions. Thus, the density has to increase in order to keep the pressure $P$ constant (in zero gravity environment). We also have from the state equation for a dilute gas, $P \propto \rho_w T_w \propto V^2 N^{\alpha - \beta}$. 

The dependence on $N$ of the collision frequency can be understood as follows: we have $\nu_w \propto \rho_w \overline {v}$. In a molecular gas, $\rho_w \propto N$ and $\overline {v}$ is fixed by the thermostat and does not depend on $N$, thus we have $\nu_w \propto N$. In a granular gas, $\overline {v}$ or the total kinetic energy $E$ are determined from the balance between the injected power by the vibrating piston and the dissipated one by inelastic collisions. For a dilute gas with a restitution coefficient $r$ very close to 1 such that the density is roughly homogeneous ($\rho_w \propto N$), $E$ does not depend on $N$, and thus $\overline {v} \propto 1/\sqrt{N}$. Thus, we get  $\nu_w \propto \sqrt{N}$ \cite{Aumaitre2}. In the present experiment, both the scaling of the density and the one of the mean velocity differ from this limit case, because gradients of density and granular temperature cannot be neglected. However, the prediction in the limit $r \simeq 1$ gives a  good approximation to the observed scaling of the collision frequency. This shows that a granular gas driven by a vibrating piston strongly differs from a molecular gas in contact with a thermostat even in the limit $r \simeq 1$. 

\acknowledgments
We thank P.~Chainais and S.~McNamara for discussions. This work has been supported by the European Space Agency and the Centre National d'\'Etudes Spatiales. The flight has been provided by Novespace. {\em Airbus A300 Z\'ero-G} aircraft is a program of CNES and ESA. We gratefully acknowledge the Novespace team for his kind technical assistance.


\end{document}